\documentstyle[12pt,aasms4]{article}
\begin{document}
\title{QUASAR RADIO STRUCTURE IN CLUSTER ENVIRONMENTS}
\author{J.B.Hutchings}
\affil{Dominion Astrophysical Observatory, National Research Council of Canada\\
5071 W. Saanich Rd., Victoria, B.C. V8X 4M6, Canada}
\authoremail{hutchings@dao.nrc.ca}
\author{A.C.Gower}
\affil{Dept of Physics and Astronomy, University of Victoria\\
P.O Box 1700, Victoria, B.C, Canada}
\authoremail{agower@otter.phys.uvic.ca}
\author{S.Ryneveld and A.Dewey}
\affil{Dominion Astrophysical Observatory, National Research Council of Canada\\
5071 W. Saanich Rd., Victoria, B.C. V8X 4M6, Canada}
\begin{abstract}
VLA\footnote{The Very Large Array is a part of the National Radio Astronomy
Observatory, which is operated by Associated Universities Inc, under
contract with the NSF}
 B-configuration snapshots have been obtained at 6cm and 20cm of 43 radio
quasars with a range of known galaxy cluster environments. The radio sources
are characterised by measures of morphology, flux, and size. These are studied
as a function of cluster density, characterised by published B$_{gq}$ numbers.
We find no strong dependence of radio source properties on this measure of
cluster density. We find $\sim2\sigma$ evidence that sources in higher cluster
densities have more lobe-dominated morphology with higher lobe luminosities and
lower core luminosities. The 7 unresolved sources are found in low density clusters,
as are the most bent sources. We find that QSO clusters have fewer other radio sources 
within a radius of 3 arcminutes than do the central galaxies of a sample of rich X-ray
selected clusters.
\end{abstract}
\section{Introduction}
    We have obtained radio images of 43 QSOs of redshift  0.7 or less with a range of
known galaxy environments from optical studies. The program was planned to study 
the connection between radio source properties and the galaxy cluster density
around the QSO.The cluster density of the
QSOs is measured by the B$_{gq}$ number derived from the galaxy companion data
largely published separately by Yee and Ellingson (1993 and references therein).
These  numbers have units of Mpc$^{1.77}$. 
We have optical images of some of the clusters, but do not yet 
have a uniform dataset that allows us to measure other cluster properties,
such as size, or the location of the QSO within the cluster.

 The radio observations were made in the B-configuration of the VLA,
in December 1991, taking $\sim$5 minute snapshots of ~50 sources at 6cm and 20cm,
interspersed with calibration sources, and determining the flux density scale
using 3C48 and 3C138. The images were self-calibrated where possible.
The synthesised beam in the final images is $\sim$4 arcsec at 20cm and $\sim$1.3 
arcsec at 6cm. The largest-scale structure fully determined in the B-configuration
observations is $\sim$100 arcsec at 20cm, and $\sim$35 arcsec at 6cm, 
and care was taken
when discussing the radio structure of QSOs with angular sizes comparable with 
or larger than these values. 
We have also used the A-configuration maps of Price et al (1993) 
and Gower and Hutchings (1984) of 10 of the sample QSOs, and also of 11 more 
QSOs with B$_{gq}$ measures, for which we have A-configuration measures by 
Hutchings, Price and Gower (1988), or Gower and Hutchings (1984).        
This gives us a total sample of 54 QSOs with radio data, of which we have
B$_{gq}$ values for 45. 
    
    A similar investigation has been published by Rector, Stocke, and Ellingson
(1995), based on published radio maps of QSOs with known B$_{gq}$ values. Since
the B$_{gq}$ value is the common parameter, there is an overlap of 16 objects
in our specifically observed sample, and another 9 for which we had earlier maps,
within their sample of 30 objects. We discuss the comparison of conclusions at the
end of this paper.

    The maps and related details will be published separately. The list and
measures relevant to this paper are listed in Table 1. The sources were 
characterised by  measures of flux density, morphology, size, and shape, 
also shown in Table 1. The bend
angles are defined by the core and hot-spots in the lobes and are usually
well-defined. The source types are defined in the Table 1 footnotes, and are
based on whether one or two lobes are present, and the relative flux densities 
of core and lobes at 20cm. The classification depends on the resolution 
of small-scale structure (hence redshift), and dynamic range where the 
lobe flux per beam area is very low. The system is the same
as that used in papers on QSO and radio galaxy structure from A-configuration maps,
as referred to above. Many of the quantities in the A-configuration maps
are taken from the measures in the references given. We have also listed a source
`complexity' measure assigned independently by two of us from visual inspection
of the maps. This is meant mainly to quantify non-symmetry which is not measured by
source type and bend.

We also noted where other faint radio sources were detected
in the field, to investigate the presence of other cluster sources, and also
to study the possible effect of the cluster on background sources.
The map dynamic range was usually in the range 500 to 1200 (peak flux density 
to RMS noise). Values outside this range were found in exceptionally weak sources 
where thermal noise dominated at 0.1 mJy at 20cm and 0.04 mJy at 6cm, or strong 
sources where dynamic range dominated ($\sim$2000 for peak fluxes above 2Jy: 
4 sources have noise above 1mJy at 6cm and 3 have noise above 2mJy at 20cm).

\section{Cluster environment}

      The distribution of the B$_{gq}$ values indicates a spread of values 
about zero, together with a distribution of higher values (see Figure~\ref{fig1}). 
From this distribution, we consider values below
350 as representing the statistical spread around the mean galaxy-galaxy correlation
amplitude - i.e. not signficantly clustered. Values above
350 are taken to represent the significantly clustered environments. 
Figure~\ref{fig2} shows the distribution of B$_{gq}$ with redshift: in order to
use a subset matched in redshift for high and low B$_{gq}$, we select QSOs
with redshift higher than 0.4. Table 2 shows mean and median values of properties 
in this matched sample of 32, divided into the high and low B$_{gq}$ groups. In
measuring
the lobe fluxes, we also rejected 2 more sources with sizes larger than 85", which
may have missing flux. (These lie one in each group, and in fact do not alter
our results. We also retain these 2 objects in the other radio source measures, 
which are not affected by missing flux.) 

     Table 2 suggests the following differences: 1) The high B$_{gq}$ group has  
a more lobe-dominated mean type (see Figure~\ref{fig3}): \it all \rm
compact sources are in low B$_{gq}$ environments). 2) High B$_{gq}$ environment
sources have higher lobe luminosities, by a factor 2-3 (see Figure~\ref{fig7}).
3) High B$_{gq}$ sources have lower core luminosities (by a factor $<$2).
There is a strong relationship between size and core luminosity (Figure~\ref{fig8}
and Neff and Hutchings 1990, first noted by Hutchings Price and Gower 1988), 
which relates this with the next point. 
4) High B$_{gq}$ sources have larger source size, particularly when 
the unresolved sources are included 
(Figure~\ref{fig4}). 5)  High B$_{gq}$ sources have less bent sources (Figure~\ref{fig5}.).

We do not quote uncertainties in Table 2 as most of the quantities do not have
gaussian distributions, and differences between properties in the two groups cannot be
quantified simply in terms of their spread. To test significance we have done 
K-S tests, which indicate that points 1 - 5 above are significant at the 2$\sigma$ 
to 3$\sigma$ level. Also, there are significant continuous correlations with 
B$_{gq}$ for the luminosities (Figure~\ref{fig7}). However, the lobe length ratio,
source complexity, and core spectral index, show no dependence on B$_{gq}$. 
    
These results are somewhat counter to the expectation that a more
dense cluster environment would bend or inhibit extended source growth. This
could indicate that in higher density clusters, the IGM is less chaotic or
dense, or that galaxy density is not related to the state of the IGM at these
redshifts. 
 
\section{Source counts}
\vskip 10pt
     In several of the B-configuration maps, one or two other sources were detected.
These are unresolved and typically faint, but all those listed in Table 1  
are 12$\sigma$ or more above their RMS map noise levels. It is of interest to 
study these extra sources to understand whether they are in the QSO cluster, or
background sources, with possible lensing by the cluster. The detection of these
sources will be affected by the map noise (usually determined by the main source
flux density), the bandwidth smearing with distance from the map centre, and the 
source flux density itself. The primary beam attenuation at 20cm 
is negligible over the
small field radii considered here. If the sources are in the cluster then the 
limiting detectable luminosity will also depend on the cluster redshift.
We look at these different effects in turn.

     Overall, a total of 22 extra sources are detected in the 43 fields in the 
20cm maps. At 20cm, the
density of sources per unit sky area around the map centre is constant out to radius
$\sim$3.2 arcmin, and then falls rapidly, with only five lying outside that.
The fall-off may begin at radius 2.6 arcmin, but by less than 30\%.
This is just where we expect the bandwidth smearing to become noticeable. Thus, we
exclude the sources beyond 3.2 arcmin. The number and flux level of extra
sources is not correlated with the cluster redshift, and so we apply no redshift
filter or correction.

 In the 6cm maps, we detect 8 extra sources, of which only one is not seen at 20cm. 
Some of the sources are marginally resolved at 6cm, but none have definite structure.
The one 6cm source not seen at 20cm may be a component of another nearby seen at both
frequencies, but otherwise there is no reason to suppose that different sources in 
a field are related to each other.

    At 20cm, the number of sources detected per field is 
about the same for all map noise levels below 0.8 mJy (Figure~\ref{fig10}).
However, the lowest flux density of detected
sources appears to rise, as expected, for noise levels above 0.3 mJy (see Figure~\ref{fig11}).
The 6cm noise levels and flux densities are lower, and 
extra sources are detected only in maps of noise 0.15 mJy and less.  
In Figure~\ref{fig11} we can see that at 20cm we detect 
6 ($<$3.2arcmin) sources in 10 eligible fields with noise $<$0.3 mJy,
and 16 sources in 25 eligible fields with noise $<$0.8 mJy. 
These fractions are very similar (0.60 and 0.64), because the low noise
fields have no sources above 10mJy and only one above 6mJy (perhaps the
result of small number statistics). 
Our result is thus that the average QSO cluster contains $\sim$0.4
extra sources within 3.2 arcmin, to a 20cm flux limit of $\sim$ 4 mJy.
To a limit of 1mJy this number is $\sim$0.6 per field.
In the 6cm maps the fraction is 0.3 per field, with a lower flux density limit,
but very small number statistics. In addition, the primary beam response varies
significantly over this field size at 6cm, so that some sources at large radii 
may fall below the detection limit.

    We looked for counterparts in the known galaxy companions to the QSOs, but 
almost all the sources lie outside the radius of the distribution of 
known companions. The few closer ones are not coincident with the (incomplete)
list of known cluster members. 
The total flux of all background sources at 20cm expected 
in this area of sky, is $\sim$17mJy, which is more than the mean extra
source flux we detect (6.1mJy average from the 25 0.8mJy-noise eligible fields). 
Thus, the extra source flux does not exceed that expected from background sources.

    As a comparison, we looked for the extra sources detected in our maps of very
rich (X-ray selected) clusters, from the CNOC program (paper in
preparation). (This progam was undertaken by the Canadian Network for Observational
Cosmology and is described e.g. by Yee, Ellingson and Carlberg 1996.)
These clusters do not have
strong central sources, but in all 5 cases, the central cD galaxy was detected
as one of the strongest sources. We thus counted extra sources around the cD
galaxies, with the same flux density and area limits as in our QSO clusters.
The rich cluster maps have lower noise and detection limits than those of the
QSO program. The numbers are summarized in Table 3.

    The comparisons in Table 3 show that the rich clusters contain on average
more sources than the QSO clusters. 
However, their total flux density does not exceed the expected background.
The rich clusters have slightly lower mean redshift than the QSO clusters. 
This has implications for source luminosity limits if they are cluster members,
which we discuss in the last section. In
the central regions of the rich  clusters, we have deep optical images, in which
we may identify the radio sources. So far, we have only looked at 
the exceptionally rich A2390 cluster, and only one of the 6 of the 
central sources corresponds with an optical
source to limits 6 magnitudes fainter than the cD galaxy (Abraham et al 1996). 

    The mean B$_{gq}$ of the CNOC clusters is 1400, compared with 300 for the QSO
sample, although some of the QSO clusters have  B$_{gq}$ near to 1000. There is
no correlation between B$_{gq}$ and number of sources in the QSO sample, so that
it may be that X-ray selection finds clusters with properties not measured by
B$_{gq}$. In any case, the preliminary evidence is that one rich X-ray selected
cluster has an enhanced population of radio sources near its centre, which 
is either associated with optically very faint objects, or represents 
background source fluxes which are enhanced, perhaps by a gravitational 
lensing effect.

\section{Other radio source properties}

     The new B-configuration data generally show the same results as our earlier
work using A-configuration maps of over 250 radio quasars of redshifts up to 3.5.
These are discussed in papers by Neff and Hutchings (1990), and Lister,
Hutchings and Gower (1994), and references therein.

      There is a correlation between core luminosity and source size 
(Figure~\ref{fig8}), but not with 
lobe luminosity. We previously discussed this in terms of source evolution in which
a strong core source fades as the lobes grow. The relation between bend angle and
size (Figure~\ref{fig9}) is similar to that of Lister et al (1994). 
There is no correlation between
bend angle and either core or lobe luminosity. The more bent sources are also
more complex on a more widespread level, and the most complex sources are
generally small. This suggests that sources which are complex have been affected
by a nuclear event or by a dense medium, or are significantly foreshortened
by projection. Finally, the most bent sources have lobes of nearly equal 
luminosity, while those with most uneven lobes are not bent. 

\section{Discussion}

   Rector et al (1995) have published tables and analysis similar to ours on the
QSO source structure. Their conclusions are in agreement with ours, but there
are some detailed differences we should note. The individual source sizes and bend
angles differ in detail, presumably as a result of map noise, dynamic range,
resolution, and occasionally differences in interpreting complex sources.
An example of the latter is 0903+169, where we give a bend of 30$^o$ compared
with their value of 58$^o$ (which they flag as uncertain). In this instance
there are several compact knots near the 20cm map centre, and the bend angle measured 
depends on which is assumed to be the core. We believe we have chosen correctly
by adopting the one which is brightest in our 6cm map. Some of their bend
measures depend on the optical position when the radio core was not visible:
our measures are based on a homogenous set of maps obtained by ourselves, and
only radio source features are used. Thus, our database is more complete and
uniform.
Similarly, the flux measures are all measured from our maps, and we regard them
as more complete and homogenous than those quoted by Rector et al.
In some cases, there are significant differences between their values and ours,
which look like cases where the core or lobe fluxes were difficult to measure or
may have suffered from poor dynamic range. We also note that the linear sizes
used by Rector et al, correspond to H$_0$=50, and not 100 as stated in their
paper. Thus, the sizes we quote (for H$_0$=100, q$_0$=0.5) are about a factor
two smaller than theirs. Finally, we note that they have chosen to split high and
low B$_{gq}$ at 500, which corresponds to a certain Abell richness, while we
have chosen 350 on the basis of the numbers in our sample. Again, the difference is
not very signficant, and a matter of choice. 

    The overall conclusion we reach is similar to that of Rector et al: that there are
no strong correlations with B$_{gq}$. The significance level of the differences
we have highlighted between high and low B$_{gq}$ sources are at the 2 to 3$\sigma$
level, either by K-S tests between the two groups, or by the correlation with 
B$_{gq}$ as a parameter.

    The lack of strong correspondence with B$_{gq}$ suggests that this quantity does
not measure the IGM well. There have been recent reports of correspondence
between radio and hot gas (X-ray) structure, in low redshift clusters such as
Virgo and Perseus (e.g. Brinkman 1996 and Owen 1996). If QSOs are at the centre 
of their clusters, they may have little movement through the cluster, and any 
IGM movement may be infall (i.e. not transverse to the radio structure). 
It would be of interest to look for radio structure correlations with 
the QSO position within the cluster, and the velocity structure of the clusters.
Ellingson, Green and Yee (1991) have published velocity studies of a few of
the QSO clusters in the sample, but there are too few measurements to make any
comparisons. They do note that  the cluster velocity dispersions are fairly low,
and that they see no significant differences in the relative velocities of companion
galaxies between radio-loud and radio-quiet QSOs. 

     We have counts of sources in the QSO cluster fields, which are lower than
those found in very rich clusters at comparable redshifts. We have not yet been
able to determine whether these extra sources are associated with cluster galaxies.
If they are cluster members, the mean difference in redshift corresponds to
a difference of a factor close to 2 in limiting luminosity. If they are background
sources, there may be implications of gravitational lensing. We are pursuing 
these studies with optical investigations. 

    Finally, we note that our new data add to and reinforce our results from
earlier studies of QSO radio structure in the distribution of bend angle and core
flux with apparent size. Thus, QSOs in cluster environments do not appear to
have large differences from the behaviour of QSOs selected without knowledge of
their environment.

    We thank Erica Ellingson and Travis Rector for useful discussions. The
new radio maps were made in collaboration with Erica Ellingson and will be
published separately.

\newpage

\centerline{Captions to figures}
\vskip 10pt
\figcaption[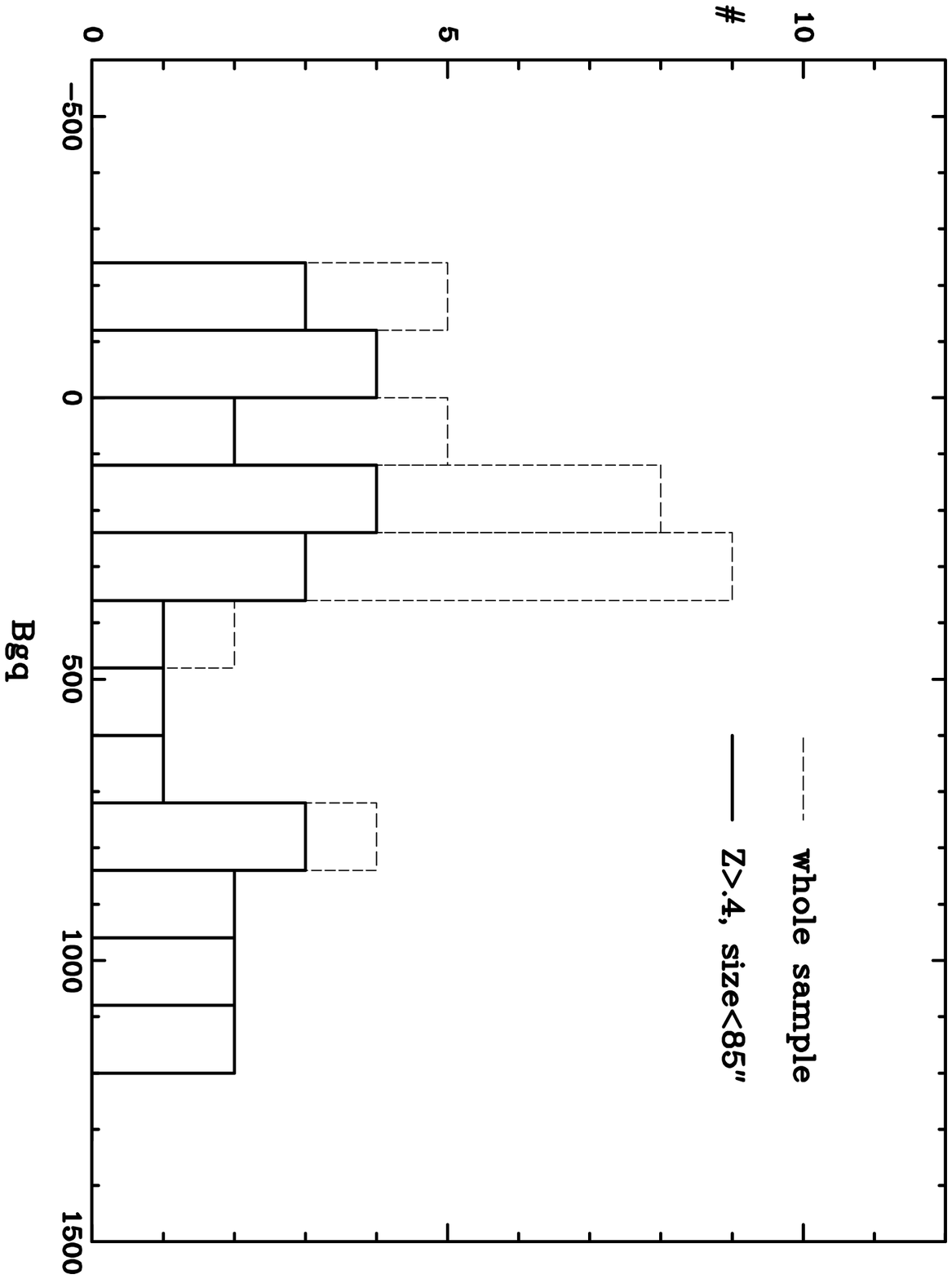]{Distribution of B$_{gq}$ values of whole sample, and
the subsample matched in redshift. \label{fig1}}

\figcaption[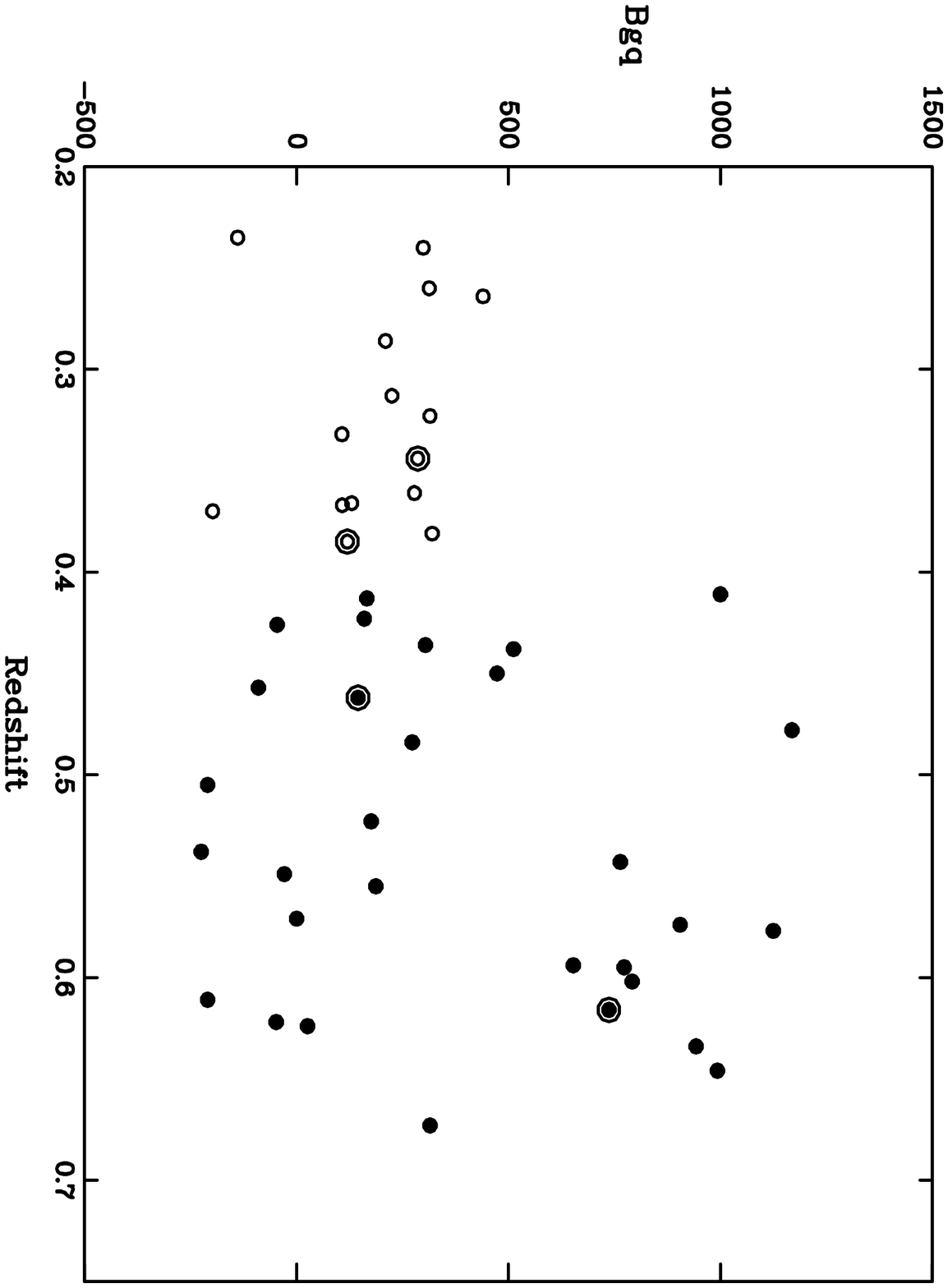]{Redshift distribution of B$_{gq}$ values. Solid symbols
are matched z$>$0.4 subset, and large symbols are sources larger than 85",
which may have missing lobe flux. \label{fig2}}

\figcaption[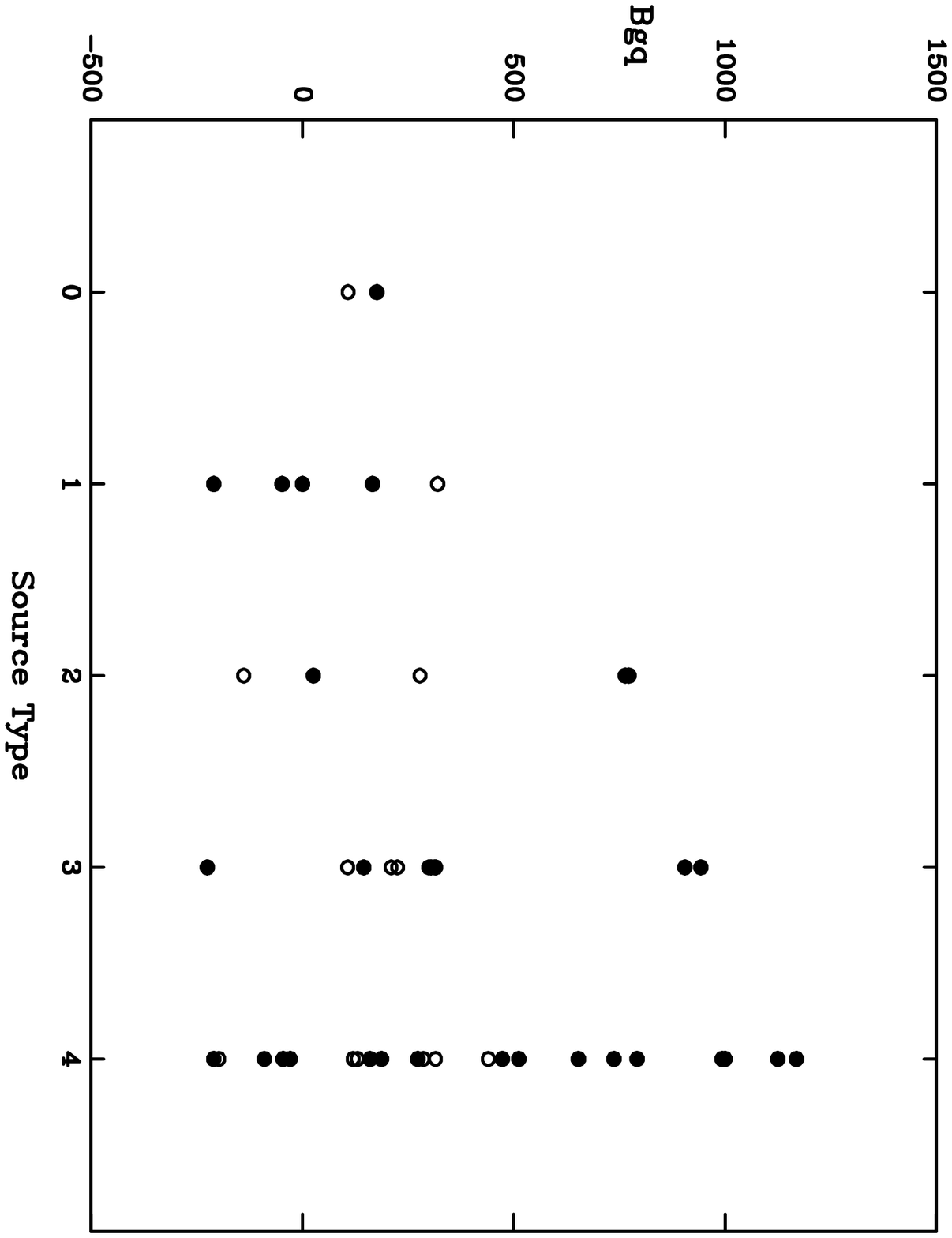]{B$_{gq}$ values as a function of source type (see Table 1
footnote). Open symbols sources with z$<$0.4. High B$_{gq}$ environments have 
no small or unresolved sources. \label{fig3}}

\figcaption[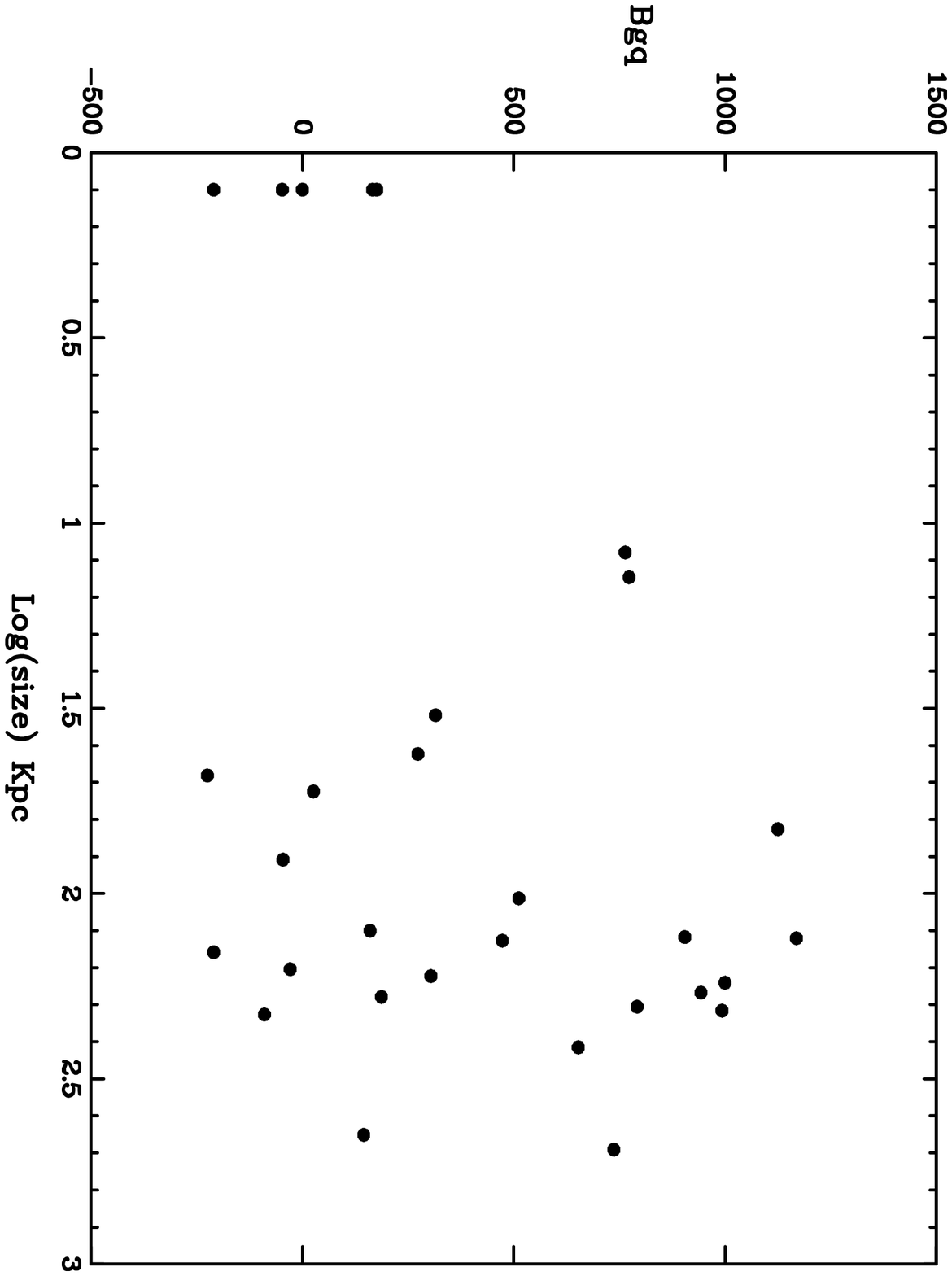]{B$_{gq}$ values for z$>$0.4 subset as a function of source
size. Five unresolved or just-resolved sources have arbitrarily low size values.
\label{fig4}}

\figcaption[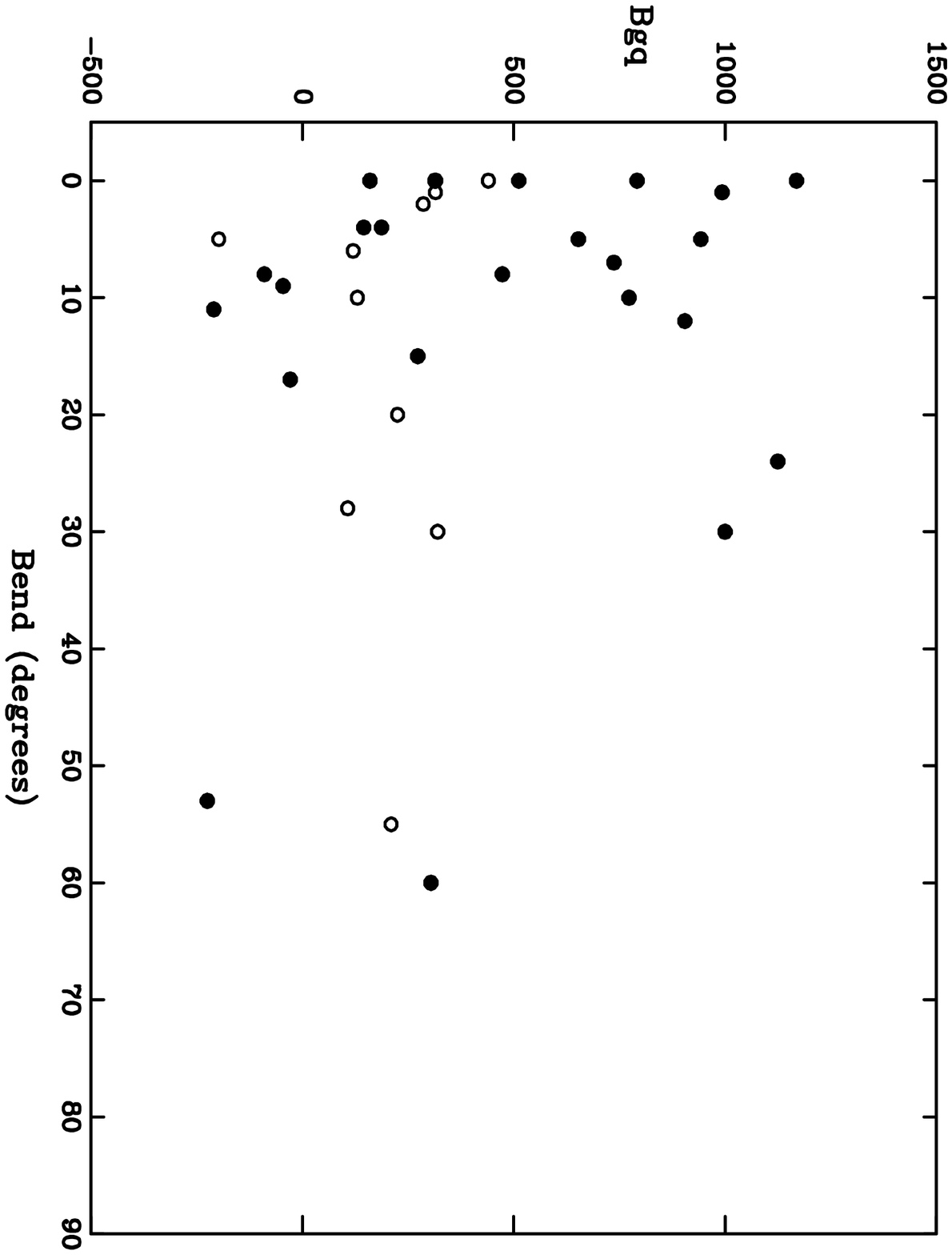]{B$_{gq}$ values with bend angle. Solid symbols are from
z$>$0.4 subset. \label{fig5}}

\figcaption[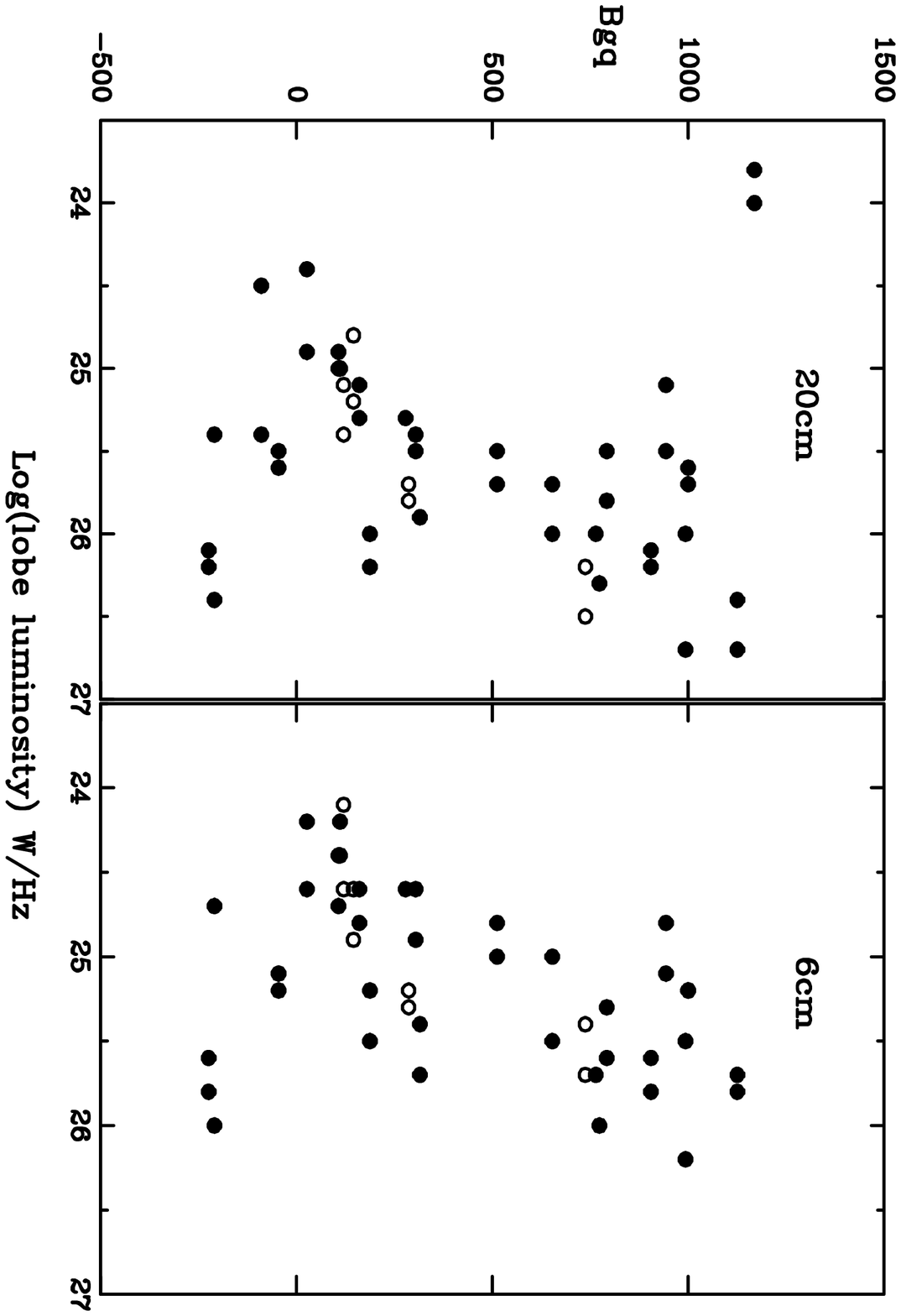]{B$_{gq}$ with lobe luminosities. Open symbols represent
four sources with sizes larger than 85", which may have missing flux. \label{fig7}}

\figcaption[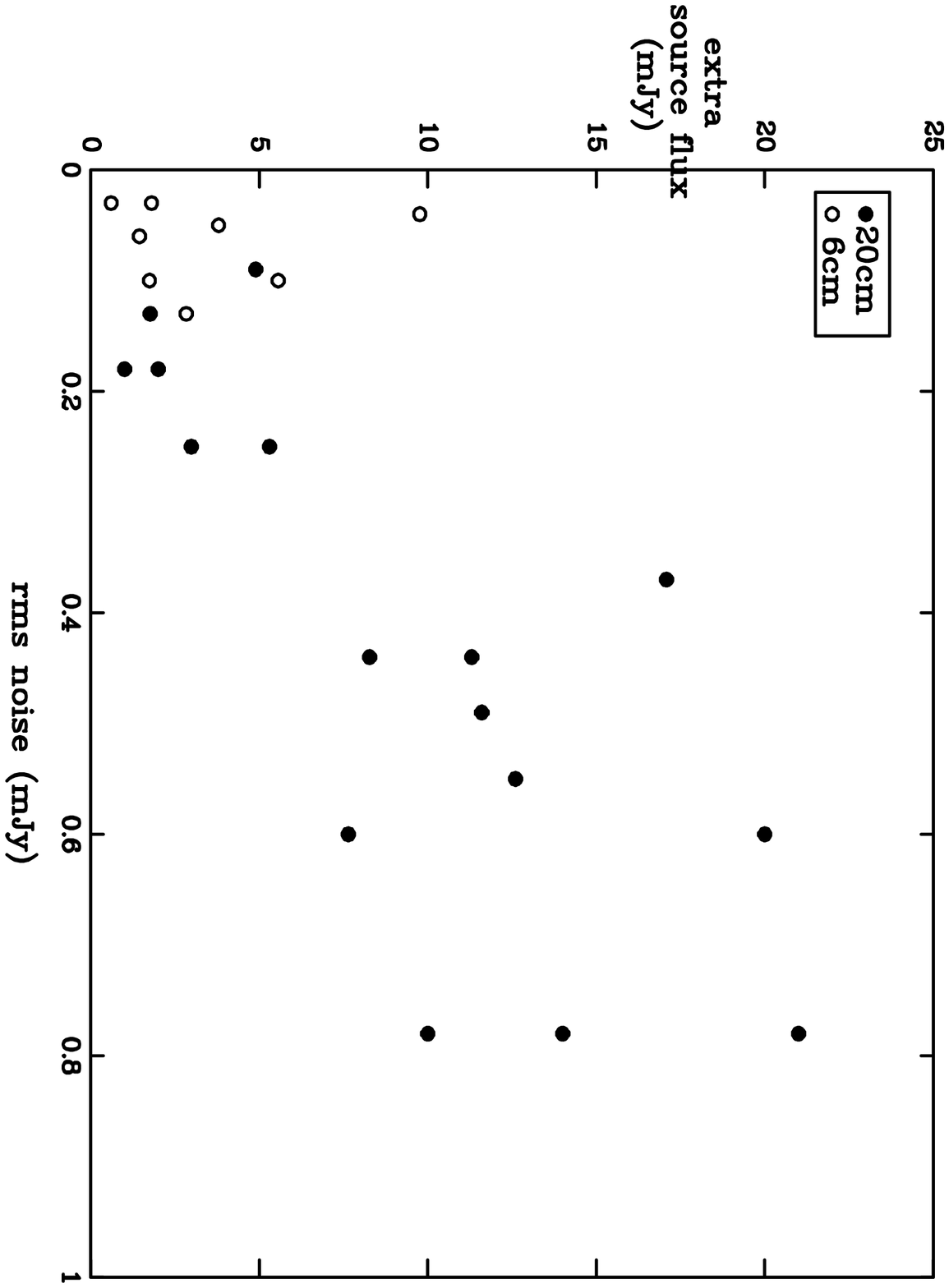]{Number of extra sources detected within 3.2 arcmin, showing
the map RMS noise lower limits of 0.8 and 0.2 mJy at 20cm and 6cm, respectively.
\label{fig10}}

\figcaption[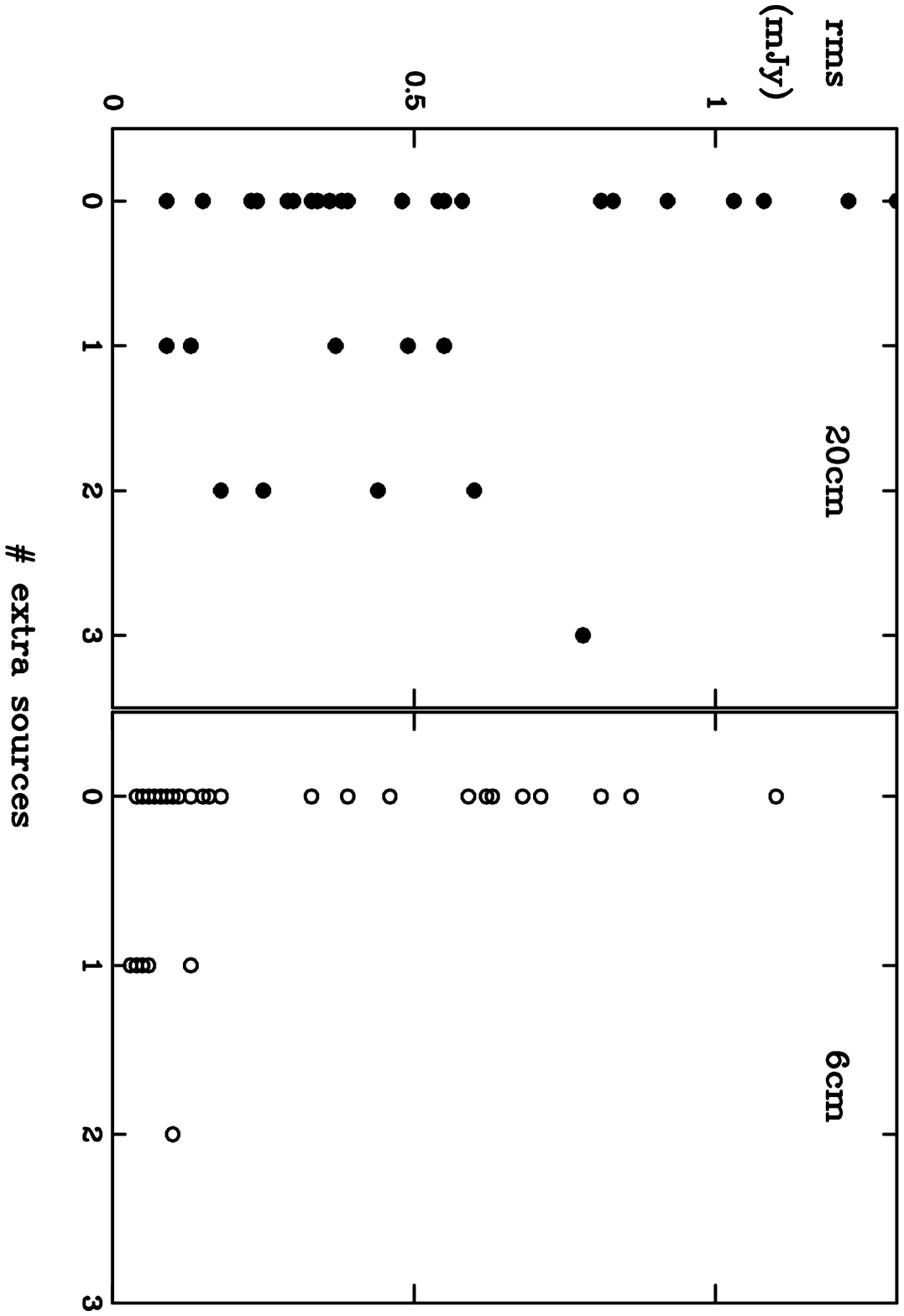]{Flux densities of extra sources as a function of map 
RMS noise.
\label{fig11}}

\figcaption[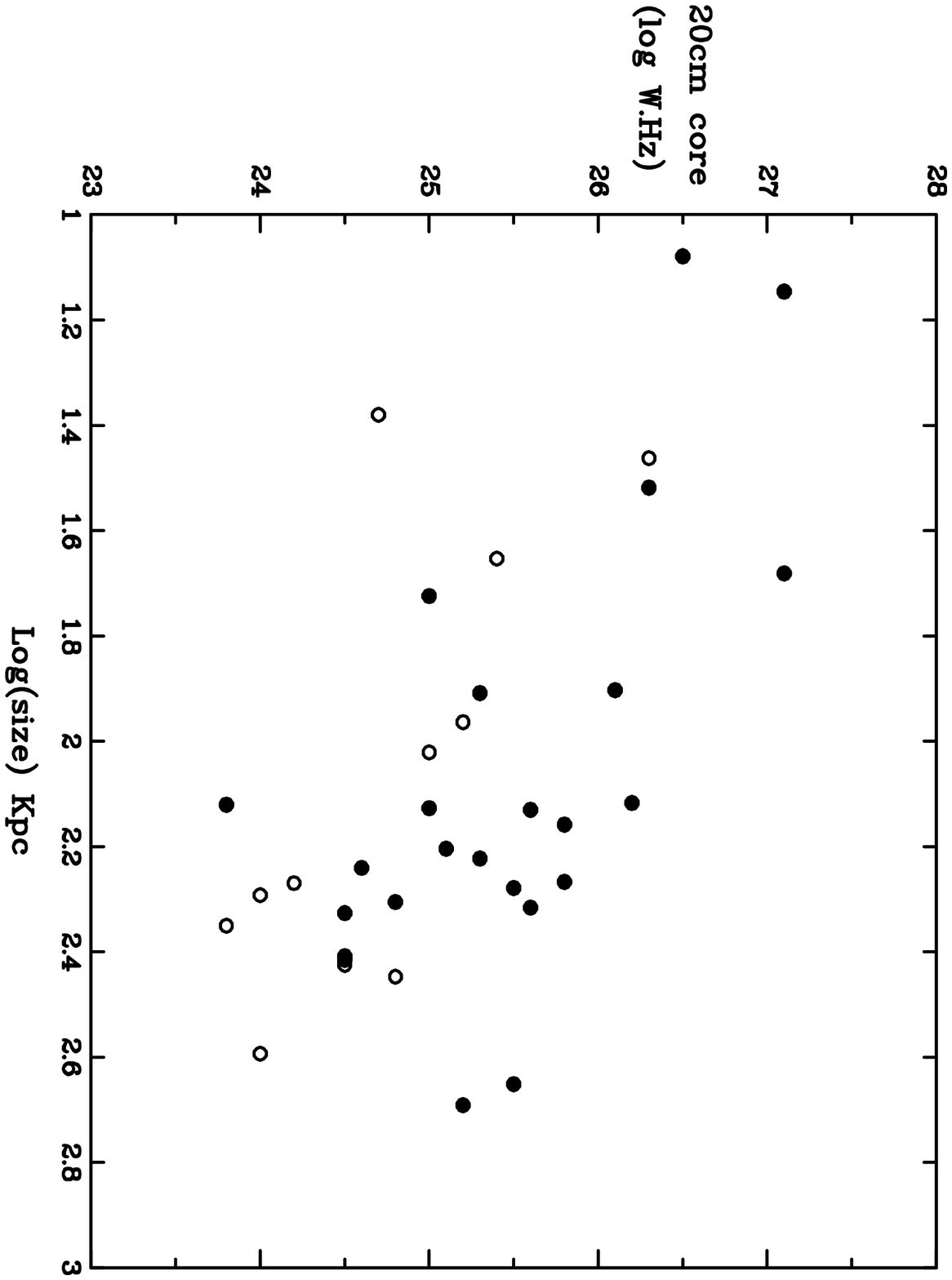]{Core luminosity with source size. Solid symbols are from
z$>$0.4 subset. \label{fig8}}

\figcaption[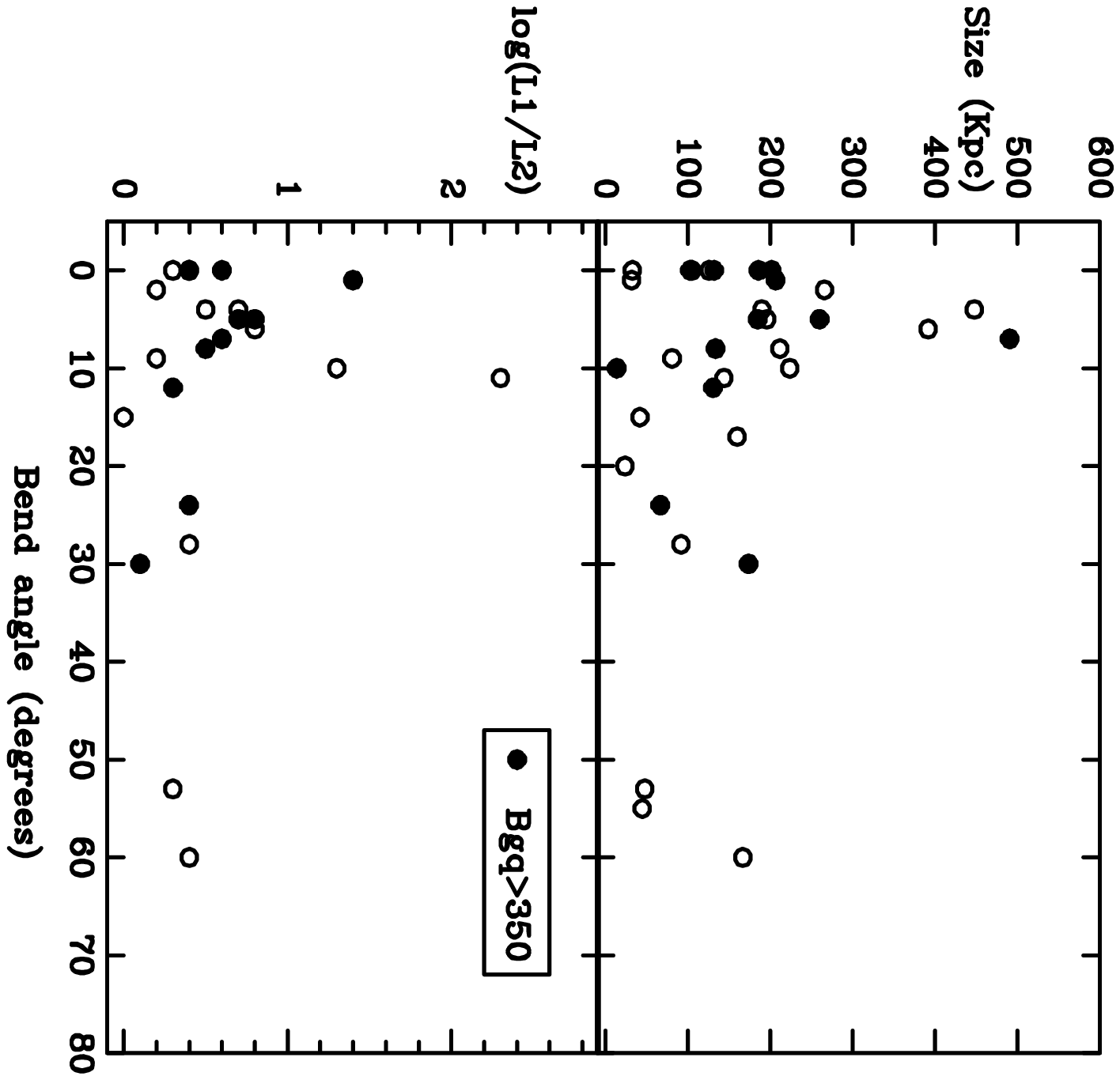]{Source size and lobe luminosity ratio with bend angle.
Trends like these are modelled with orientation by Lister et al (1994).
\label{fig9}}

\clearpage
 
\end{document}